\documentclass{aa}
\usepackage{txfonts}
\usepackage{psfig}

\begin{document}


\title{Evaluating GAIA performances on eclipsing binaries.}
\subtitle{II. Orbits and stellar parameters for V781~Tau, UV~Leo and GK~Dra}

\author{
       T. Zwitter\inst{1}
\and   U. Munari\inst{2,3}
\and   P.M. Marrese\inst{2,4}
\and   A. Pr\v sa\inst{1}
\and   E.F. Milone\inst{5}
\and   F. Boschi\inst{2}
\and   T. Tomov\inst{6}
\and   A. Siviero\inst{2}
       }
\offprints{T. Zwitter}

\institute {
University of Ljubljana, Department of Physics, Jadranska 19, 1000 Ljubljana, Slovenia 
\and
Osservatorio Astronomico di Padova, Sede di Asiago, I-36012 Asiago (VI), Italy
\and 
CISAS, Centro Interdipartimentale Studi ed Attivit\`a Spaziali dell'Universit\`a di Padova, Italy
\and
Dipartimento di Astronomia dell'Universit\`a di Padova, Osservatorio Astrofisico, I-36012 Asiago (VI), Italy
\and
Physics and Astronomy Department, University of Calgary, Calgary T2N 1N4, Canada
\and
Centre for Astronomy, Nicholaus Copernicus University, ul. Gagarina 11, 87-100 Torun, Poland
}
\date{Received date..............; accepted date................}

\abstract{
The orbits and physical parameters of three close, double-lined G0
eclipsing binaries have been derived combining $H_P, V_T, B_T$ photometry
from the Hipparcos/Tycho mission with 8480-8740 \AA\ ground-based spectroscopy. 
The setup is mimicking the photometric and spectroscopic observations that
should be obtained by GAIA.
The binaries considered here are all of G0 spectral type, but each with its 
own complications: V781 Tau is an overcontact system with components of 
unequal temperature, UV Leo shows occasional surface spots and GK Dra 
contains a $\delta$~Scuti variable. Such peculiarities will be common among 
binaries to be discovered by GAIA. We find that the values of 
masses, radii and temperatures for such stars can be 
derived with a 1-2\%\ accuracy using the adopted GAIA-like observing mode. 

\keywords{surveys:GAIA -- stars:fundamental parameters -- binaries:eclipsing -- 
binaries:spectroscopic}
}
\maketitle

\section{Introduction}

GAIA is a challenging Cornerstone mission re-approved by ESA last May for a
lunch by around 2010. It is aimed to provide micro-arcsec astrometry, 10-band
photometry and medium resolution 8480--8740\AA\ spectroscopy for a huge number of
stars, with completeness limits for astrometry and photometry set to $V=20$
mag. Each target star will be measured around a hundred times during the
five year mission life-time, in a fashion similar to the highly successful
operational mode of {\sl Hipparcos}. The astrophysical and technical guidelines
of the mission are described in the ESA's {\sl Concept and Technology Study}
(ESA SP-2000-4), in the papers by Gilmore et al.\ (1998) and Perryman et al.\ 
(2001), and in the proceedings of conferences devoted to GAIA and edited by
Strai\v{z}ys (1999), Bienaym\'{e} and Turon (2002), Vansevi\v{c}ius et al.\ 
(2002) and Munari (2003).

In Paper~I of this series, Munari et al. (2001), we have started to 
provide reasonable orbits for a number of new eclipsing binaries and to 
evaluate expected performances of GAIA on eclipsing binaries with an emphasis on
the achievable accuracy of derived fundamental stellar parameters
like masses and radii. The expected number of eclipsing binaries to be 
discovered by GAIA is $\sim4 \times 10^5$. Some $10^5$ of these will be 
characterized as double-lined in GAIA spectral observations. This is a 
huge number, many orders of magnitude larger than the total of SB2 
eclipsing binaries so far investigated from ground-based observations (cf.\ 
Andersen 1991; Batten, Fletcher \&\ MacCarthy 1989). Perhaps the orbits and stellar parameters could be
derived from GAIA observations at a few percent error only for a few percent 
of them. But this still represents a two-orders of magnitude increase compared 
to all ground-based observing campaigns during the last century. Data obtained 
by GAIA should be able to provide reasonable solutions as ground-based 
follow-up campaigns will be very time consuming. It is therefore of 
great interest to investigate the expected performances of GAIA on
eclipsing binaries. The purpose of this series of papers 
is to contribute to the fine tuning of 
the last details in the mission planning as well as to define the 
strategy to analyze the massive spectroscopic and photometric data flow 
on eclipsing binaries that is completely unprecedented.  In the meantime,
this series of papers will focus on eclipsing binaries unknown or poorly
studied in the literature so far.

\begin{table*}[!Ht]
\tabcolsep 0.08truecm
\caption{Programme eclipsing binaries. Data from the Hipparcos Catalogue. 
               $H_P, B_T, V_T$ are out-of-eclipse median values.}
\begin{tabular}{lccccccccrclccrclrcl} \hline
&&&&&&&&&&&&&&\\
Name & & Spct. &~ $H_P$ &~ $B_T$ &~ $V_T$ &~ $\alpha_{J2000}$ &~ $\delta_{J2000}$ &\multicolumn{3}{c}{parallax}&~~~dist&\multicolumn{3}{c}{$\mu_\alpha$}&\multicolumn{3}{c}{~~$\mu_\delta$}\\ 
     & &       &        &        &        &~ (h m s)          &~ ($^\circ$ ' ") &\multicolumn{3}{c}{(mas)}&~~~(pc)&\multicolumn{3}{c}{(mas yr$^{-1}$)}&\multicolumn{3}{c}{~~(mas yr$^{-1}$)}\\
&&&&&&&&&&&&&&&&&&\\ \hline
&&&&&&&&&&&&&&&&&&\\
V781 Tau & HIP 27562  & G0 &~ 8.71  &~ 9.41 &~ 8.74 &~ 05 50 13.12 &~ +26 57 43.4 &~ 12.31&$\pm$&1.35 &~~  81$^{91}_{73}$   &~ --0.084&$\pm$&0.004 &~  --0.091&$\pm$&0.004 \\
UV Leo   & HIP 52066  & G0 &~ 9.20  &~ 9.78 &~ 9.00 &~ 10 38 20.77 &~ +14 16 03.7 &~ 10.85&$\pm$&1.16 &~~  91$^{103}_{83}$  &~ --0.007&$\pm$&0.004 &~~   0.010&$\pm$&0.004 \\
GK Dra   & HIP 82056  & G0 &~ 8.92  &~ 9.19 &~ 8.81 &~ 16 45 41.19 &~ +68 15 30.9 &~~ 3.37&$\pm$&0.69 &~~ 297$^{373}_{246}$ &~ --0.013&$\pm$&0.002 &~~   0.014&$\pm$&0.002 \\
&&&&&&&&&&&&&&&&\\
\hline
\end{tabular}
\end{table*}

Paper~I outlines the framework of the project and adopted methods, and
the reader is referred to it (and the references therein) for details. In short,
Hipparcos/Tycho photometry is adopted as a fair simulation of typical GAIA
photometric data. The satellite spectroscopic data is simulated by
devoted ground-based observations obtained with the Asiago 1.82m + Echelle +
CCD, set up to mimic the expected GAIA spectra. Precision of the results 
of our investigation can be considered as a lower limit to the accuracy 
obtainable from GAIA
at the given source S/N, because ($a$) GAIA will observe in many more
photometric bands than Hipparcos/Tycho and with far higher accuracy even 
in the narrow bands, thus both increasing light-curve
mapping as well as accuracy of information on stellar temperature,
limb-darkening and reddening; and ($b$) GAIA will acquire at least twice as many 
spectra per star than considered here due to obvious limitations in the 
telescope access time. 

\section{Target selection}

Similar to Paper~I we have selected both some brand-new eclipsing binaries (i.e.\ 
without a spectroscopic or photometric orbit solution in the literature) as
well as binaries with already published orbital solutions (however not in
the GAIA spectral range) that can serve as an external comparison. Their 
basic properties are quoted in Table~1.

{\it V781~Tau}. This is a G0 over-contact ($\sim$23\%) binary 
($P \sim 0.4$ days) with stars of unequal temperature. It is known to 
undergo period changes (Donato et al.\ 2003, in preparation), 
interpreted by Liu and Yang (2000) as shrinkage of the secondary. 
A spectrophotometric orbit of moderate quality has been published by 
Lu (1993).

{\it UV~Leo}. This is a G0 short period binary ($P=0.6$ days) showing
intrinsic variations caused by cool spots on the secondary component (cf.\ 
Miku\v{z} et al.\ 2002). Orbital parameters have been derived from {\it UBV}
photometric data by Frederik \&\ Etzel (1996) and from 4430-6800 \AA\
spectroscopic observations by Popper (1997).

{\it GK~Dra}. This is a newly discovered eclipsing binary, the only existing
information in the literature being $BV$ photometric monitoring by Dallaporta
et al.\ (2002). The authors showed that the photometric period listed in the
Hipparcos Catalogue ($\sim 17$ days) is wrong (the actual one being 9.97 days),
and that the secondary star has intrinsic variability of a $\delta-$Sct type.

\begin{table}[!Hb]
\tabcolsep 0.08truecm
\caption{Number of Hipparcos ($H_P$) and Tycho ($B_T$, $V_T$) photometric data 
and ground based radial velocity observations, their mean S/N and standard error
for the three programme stars. Error for radial velocity is in km~s$^{-1}$.}
\begin{tabular}{lccccccccccc} \hline
&&
\multicolumn{2}{c}{\sl Hip}&&
\multicolumn{3}{c}{\sl Tyc}&&
\multicolumn{3}{c}{\sl RV}\\ \cline{3-4} \cline{6-8} \cline{10-12} 
\multicolumn{11}{c}{}\\&&

N&$\sigma$($H_P$)&&
N&$\sigma$($B_T$)&$\sigma$($V_T$)&&
N&S/N&$\sigma$({\sl RV})\\
\multicolumn{11}{c}{}\\
V781 Tau &&  61 & 0.014 &&  81 & 0.18 & 0.15 && 41 & 35 &  8 \\
UV Leo   &&  96 & 0.015 && 150 & 0.21 & 0.17 && 29 & 30 & 10 \\
GK Dra   && 124 & 0.017 && 179 & 0.15 & 0.15 && 35 & 45 &  3 \\  
\hline
\end{tabular}
\end{table}

\begin{table*}[!Ht]
\tabcolsep 0.08truecm
\caption{Journal of radial velocity data. The columns give the spectrum
number (as from the Asiago 1.82 m Echelle+CCD log book), the heliocentric JD, 
and the heliocentric radial velocities (in km~s$^{-1}$)  
for both components. An asterisk marks the spectra with a too severe blending 
of the lines for a meaningful measurement of radial velocities of each 
component. The latter have not been used in modeling of the binaries.}
\begin{tabular}{llrrcllrrcllrr}
\hline 
\multicolumn{4}{c}{V781 Tau}&&\multicolumn{4}{c}{UV Leo}&&\multicolumn{4}{c}{GK Dra}\\
\multicolumn{1}{c}{\#}&\multicolumn{1}{c}{HJD}&\multicolumn{1}{c}{RV$_1$}&\multicolumn{1}{c}{RV$_2$}&&
\multicolumn{1}{c}{\#}&\multicolumn{1}{c}{HJD}&\multicolumn{1}{c}{RV$_1$}&\multicolumn{1}{c}{RV$_2$}&&
\multicolumn{1}{c}{\#}&\multicolumn{1}{c}{HJD}&\multicolumn{1}{c}{RV$_1$}&\multicolumn{1}{c}{RV$_2$}
\\
\multicolumn{4}{c}{------------------------------------------------}&\hspace*{3mm}&
\multicolumn{4}{c}{------------------------------------------------}&\hspace*{3mm}&
\multicolumn{4}{c}{------------------------------------------------}\\
30731* &2451153.53313&   36.4&    36.4 &&31837 &  2451209.51624&--151.4&	199.7&&31848 & 2451209.60066&--71.9&	   59.9    \\
30788  &2451154.52849&  253.0&  --66.6 &&31878 &  2451210.43364 & 166.6&      --172.8&&32100 & 2451225.52478 & 74.2&	 --59.3   \\
30802* &2451154.61290&   23.9&    23.9 &&31952 &  2451216.49289 & 116.0&       --73.3&&32775 & 2451274.54782 & 80.3&	 --63.4   \\
30852  &2451155.49988&  207.5&  --15.8 &&31968*&  2451217.46653 &  18.9&	 18.9&&32817 & 2451275.52285 & 76.5&	 --50.3   \\
30867* &2451155.63872&   28.9&    28.9 &&32012 &  2451221.51594&--156.3&	171.3&&32869 & 2451279.53257&--70.8&	   67.5   \\
30913  &2451156.52626&  160.1&  --61.6 &&32085 &  2451225.43269 & 179.4&      --158.2&&32960 & 2451339.41309&--72.3&	   63.2   \\
31171* &2451165.47882&   53.4&    53.4 &&32658 &  2451269.44151&--110.2&	128.3&&33115 & 2451402.34909 & 27.5&	 --11.0   \\
31229* &2451166.51393&   49.1&    49.1 &&32663 &  2451269.47415&--149.6&	175.1&&33978 & 2451564.59894 & 76.9&	 --60.0   \\
31278  &2451167.48421&--120.6&    88.9 &&32668 &  2451269.50678&--164.4&	206.5&&34153 & 2451589.57475&--79.3&	   63.3   \\
31327* &2451169.58493&    8.0&    08.0 &&32802 &  2451275.44879&--121.7&	133.1&&34182 & 2451592.56867 & 50.8&	 --39.5   \\
31460* &2451197.50043&  --0.3&   --0.3 &&32807 &  2451275.48073&--147.3&	183.3&&34226 & 2451593.57209 & 85.7&	 --63.9   \\
31462* &2451197.51640&   11.8&    11.8 &&33967 &  2451564.53178 &  99.9&       --98.0&&34382 & 2451621.43916 &  5.1&	    5.1   \\
31622  &2451201.26374&--176.8&   116.1 &&34228 &  2451593.60814&--112.1&	137.0&&34418 & 2451624.51494 & 72.6&	 --54.4   \\
31624  &2451201.28454&--152.1&   104.3 &&34410 &  2451624.41908 & 159.9&      --142.5&&34453 & 2451625.49986 & 46.8&	 --33.2   \\
31626* &2451201.30692&    8.7&     8.7 &&34413 &  2451624.46813 & 150.9&      --177.8&&34503 & 2451626.51482 &  4.5&	    4.5   \\
31628* &2451201.32782&   31.5&    31.5 &&34416 &  2451624.49141 & 157.4&      --128.5&&35762 & 2451798.51275 &	   &5.2   \\
31630* &2451201.34668&   46.9&    46.9 &&34443 &  2451625.42998&--102.6&	141.3&&36093 & 2451895.65844 &--5.0&	  --5.0   \\
31632  &2451201.36542&  173.6&  --22.8 &&34501 &  2451626.49333 &--98.4&	153.7&&36143 & 2451896.68512&--34.2&	   33.4   \\
31634  &2451201.38402&  228.4&  --52.1 &&36082*&  2451895.53608& --31.8&       --31.8&&36172 & 2451923.62582 & 79.5&	 --56.2   \\
31636  &2451201.40272&  254.7&  --61.8 &&36084*&  2451895.55506 &  54.9&       --54.3&&36286 & 2451924.63312 & 48.1&	 --35.8   \\
31638  &2451201.42120&  265.0&  --66.9 &&36087 &  2451895.60196 &  79.4&       --91.6&&36413 & 2451951.71656 & 55.3&	 --44.3   \\
31640  &2451201.44007&  239.2&  --53.0 &&36089 &  2451895.62120 & 106.4&      --106.8&&36437 & 2451952.52896 & 82.0&	 --62.2   \\
31642  &2451201.45878&  191.9&  --37.0 &&36095 &  2451895.68774 & 153.0&      --152.5&&36501 & 2451954.61109 & 46.6&	 --30.0   \\
31644* &2451201.48908&   42.4&    42.4 &&36133 &  2451896.56836&--151.1&	148.9&&36533 & 2451955.60739 &--3.8&	  --3.8   \\
31646* &2451201.50783&   31.5&    31.5 &&36135 &  2451896.58704&--169.7&	155.1&&36558 & 2451983.57775 & 74.2&	 --55.2   \\
31648* &2451201.52638&   17.7&    17.7 &&36140 &  2451896.64222&--134.7&	138.5&&36811 & 2452067.40354&--72.7&	   59.5   \\
31650  &2451201.54527&--143.3&    99.8 &&36142 &  2451896.66113&--136.5&	117.1&&37930 & 2452300.54093 & 44.5&	 --31.4   \\
31652  &2451201.56386&--169.6&   120.3 &&36278 &  2451924.52603 & 141.9&      --178.4&&37955 & 2452302.58732 & 77.8&	 --64.0   \\
31654  &2451201.58275&--187.5&   128.3 &&36386 &  2451951.51048 & 156.3&      --199.1&&38392 & 2452361.47712 & 82.8&	 --63.4   \\
31667  &2451202.28749&--188.7&   119.7 &&&& &					     &&38394 & 2452361.50350 & 81.8&	 --61.8   \\
31682  &2451202.46289&  265.5&  --76.4 &&&& &					     &&38518 & 2452387.46484&--80.4&	   69.2   \\
34483  &2451626.31285&  227.7&  --51.6 &&&& &					     &&38536 & 2452388.48135&--56.8&	   46.4   \\
34485  &2451626.33501&  259.7&  --65.3 &&&& &					     &&38543 & 2452389.49614 &  0.0&	    0.0   \\
34487  &2451626.35703&  248.8&  --61.0 &&&& &					     &&38561 & 2452447.46833&--77.9&	   64.9   \\
37488  &2452242.53303&--181.8&   121.1 &&&& &					     &&38579 & 2452448.36282&--53.6&	   46.9   \\
37497  &2452242.70637&  271.2&  --67.4 & &&&&&&&&&\\
37601  &2452271.35705&  184.5&  --71.5 & &&&&&&&&&\\
37627  &2452272.38422&  207.0&  --61.1 & &&&&&&&&&\\
37680* &2452277.40348&   15.7&    15.7 & &&&&&&&&&\\
37821  &2452280.46476&--191.9&   133.3 & &&&&&&&&&\\
38165  &2452330.49815&--176.4&   112.0 & &&&&&&&&&\\
\hline
\end{tabular}
\end{table*}

\begin{table*}[!Ht]
\caption[]{Modeling solutions. The uncertainties are formal mean 
standard errors to the solution. The last four rows give  the standard
deviation of the observed points  from the derived orbital solution.} 
\begin{tabular}{lrclrclrcl}
\hline
&&&&&&&&&\\
parameter(units)&\multicolumn{3}{c}{V781 Tau}&\multicolumn{3}{c}{UV Leo}&\multicolumn{3}{c}{GK Dra}\\ 
&&&&&&&&&\\
\hline
Period (days)   &0.34490857    &$\pm$&0.0000001
                                            & 0.600086 &$\pm$&0.000001 & 9.9742	 &$\pm$&0.0002\\
Epoch (HJD)     &2447962.46572 &$\pm$&0.00016&2448500.560&$\pm$&0.001& 2452005.56 &$\pm$&0.03  \\
$a$ ($\mathrm{R}_\odot$)
                &2.4478        &$\pm$&0.002 & 3.957    &$\pm$&0.087 & 28.92 	 &$\pm$&  0.35\\
V$_\gamma$ (km s$^{-1}$)
                &30.44         &$\pm$&0.10  & 3.9      &$\pm$&3.1   & 1.68	 &$\pm$& 0.67 \\
$q = \frac{m_2}{m_1}$
                &2.278       &$\pm$&0.028 & 0.917    &$\pm$&0.027 & 1.244  	 &$\pm$& 0.020\\
$i$ (deg)       &66.80         &$\pm$&1.04  & 83.07    &$\pm$&0.91  & 86.07 	 &$\pm$& 0.18  \\
$e$             &0.0           &     &      & 0.0      &     &      & 0.084  	 &$\pm$& 0.013\\
$\omega$ (deg from $a$)  
                &              &     &      &          &     &      & 175.4 	 &$\pm$& 1.4   \\
T$_1$ (K)       &6390          &$\pm$&11    & 6129     &$\pm$&67    & 7100	 &$\pm$&70   \\
T$_2$ (K)       &6167          &$\pm$&10    & 5741     &$\pm$&59    & 6878	 &$\pm$&57   \\
$\Omega_1$      &5.640         &$\pm$&0.05  & 5.024    &$\pm$&0.090 & 12.26  	 &$\pm$&0.21  \\
$\Omega_2$      &5.640         &$\pm$&0.05  & 4.093    &$\pm$&0.074 & 13.69  	 &$\pm$&0.24  \\
$R_1$ (R$_\odot$)
                &0.759         &$\pm$&0.007 & 0.973    &$\pm$&0.024 & 2.431	 &$\pm$&0.042  \\
$R_2$ (R$_\odot$)
                &1.111         &$\pm$&0.007 & 1.216    &$\pm$&0.043 & 2.830	 &$\pm$&0.054  \\
$M_1$ (M$_\odot$)
                &0.510         &$\pm$&0.006 & 1.210    &$\pm$&0.097 & 1.460  	 &$\pm$&0.066  \\
$M_2$ (M$_\odot$)
                &1.150         &$\pm$&0.027 & 1.110    &$\pm$&0.100 & 1.810 	 &$\pm$&0.109  \\
M$_{bol,1}$     &4.950         &$\pm$&0.025 & 4.590    &$\pm$&0.094 & 1.960 	 &$\pm$&0.075  \\
M$_{bol,2}$     &4.280         &$\pm$&0.020 & 4.390    &$\pm$&0.113 & 1.770 	 &$\pm$&0.072  \\
$\log g_1$ (cgs)&4.380         &$\pm$&0.012 & 4.540    &$\pm$&0.053 & 3.830 	 &$\pm$&0.033  \\
$\log g_2$ (cgs)&4.410         &$\pm$&0.016 & 4.310    &$\pm$&0.055 & 3.790 	 &$\pm$&0.041  \\
&&&&&&&&& \\
$\sigma_{RV,1,2}$ (km s$^{-1}$)
                &13.8          &     &      & 17.6     &     &      & 2.71	 &     &      \\
$\sigma (B_t)$ (mag)
                &0.193         &     &      & 0.228    &     &      & 0.187	 &     &      \\
$\sigma (V_t)$ (mag)
                &0.173         &     &      & 0.227    &     &      & 0.199	 &     &      \\
$\sigma (H_p)$ (mag)
                &0.020         &     &      & 0.028    &     &      & 0.028	 &     &      \\
&&&&&&&&& \\
\hline
\end{tabular}     
\end{table*}

\begin{figure}[t]
\centerline{\psfig{file=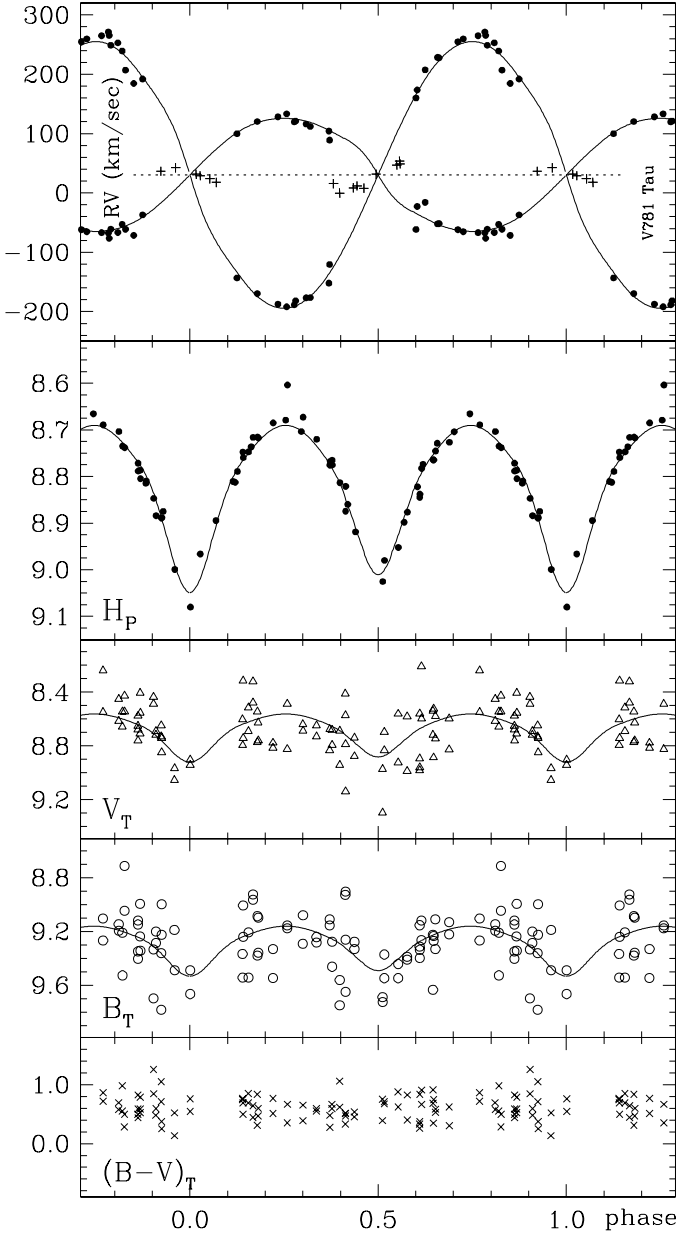,width=8.9cm}}
\caption[]{Hipparcos $H_P$ and Tycho $V_T, B_T, (B-V)_T$ lightcurves of V781~Tau
folded onto the period $P = 0.34490857$~days. 
Radial velocity measurements in the GAIA spectral interval from Table~3 
are given on the top, with '$+$' signs marking blended spectral lines that were 
not used for modeling.  
The curves represent the solution given in Table~4.}
\end{figure}

\begin{figure}[t]
\centerline{\psfig{file=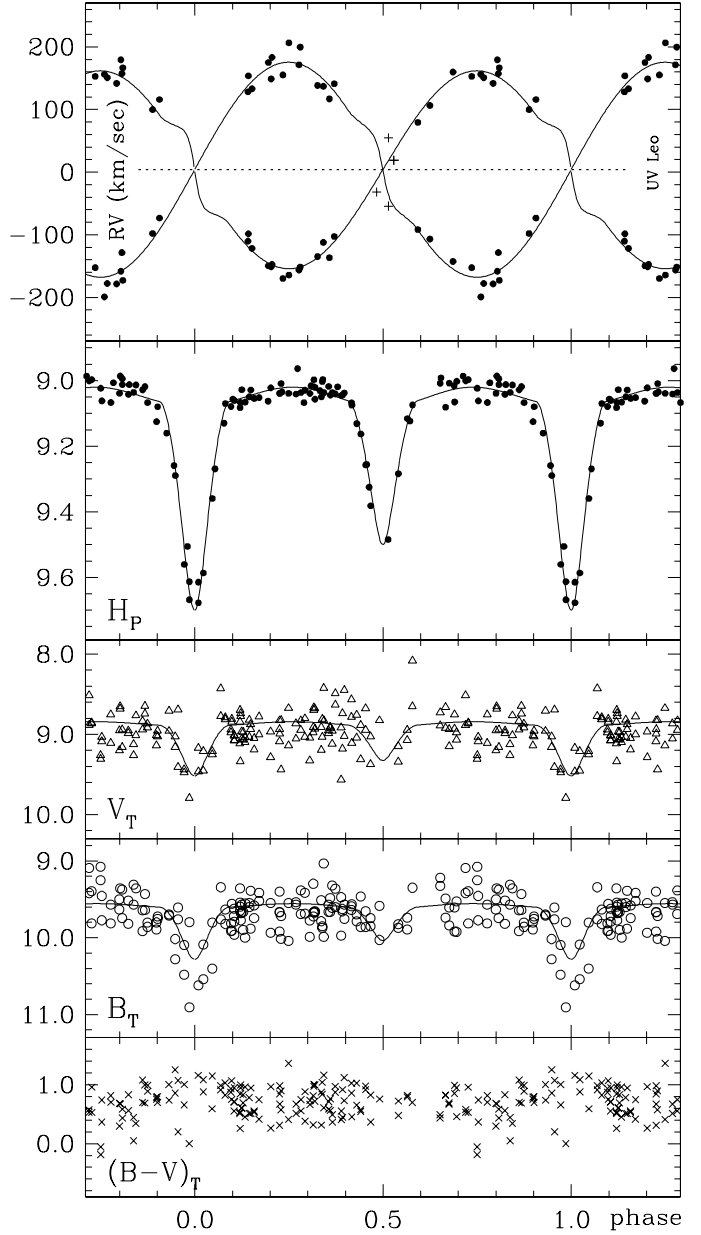,width=8.9cm}}
\caption[]{Hipparcos $H_P$ and Tycho $V_T, B_T, (B-V)_T$ lightcurves of UV~Leo
folded onto the period $P = 0.600086$~days. 
Radial velocity measurements in the GAIA spectral interval from Table~3 
are given on the top, with '$+$' signs marking measurements around primary eclipse 
that were not considered in modeling.   
The curves represent the solution given in Table~4.}
\end{figure}

\begin{figure}[t]
\centerline{\psfig{file=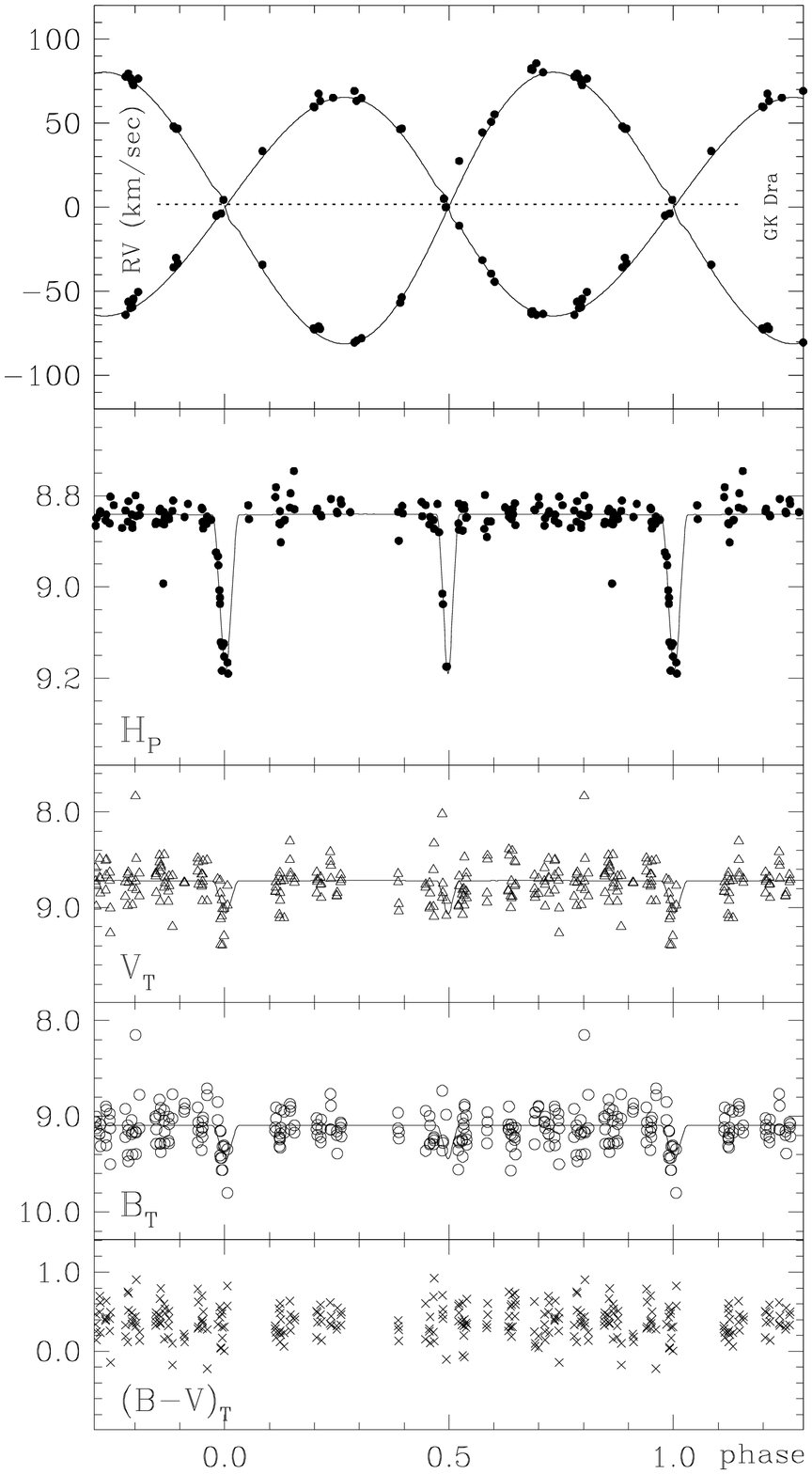,width=8.9cm}}
\caption[]{Hipparcos $H_P$ and Tycho $V_T, B_T, (B-V)_T$ lightcurves of GK~Dra
folded onto the period $P = 9.9742$~days. 
Radial velocity measurements in the GAIA spectral interval from Table~3 
are given on the top.   
The curves represent the solution given in Table~4.}
\end{figure}

\section{Observations}

As explained above we use Hipparcos photometry as a lower limit 
to the photometric information expected from GAIA. The accuracy of 
Hipparcos photometry is lower, but the number of observations of each 
star with only a limited number of points sampling the eclipses is 
similar. Table~2 gives details on the number of observations of each 
star and their accuracy.

All spectral observations were obtained in the same mode as in Paper~I, 
i.e.\ at 0.25 \AA/pix dispersion and $\sim$0.50 \AA\ resolution 
over the 8480-8740 \AA\ wavelength range (therefore a resolving power
R~=~$\lambda/\bigtriangleup\lambda$~=~17,000).

The spectroscopic observations have been collected with the Echelle+CCD
spectrograph on the 1.82 m telescope operated by Osservatorio Astronomico di
Padova atop of Mt.\ Ekar (Asiago). A 2.2 arcsec slit width was adopted to meet 
the resolution requirement. The detector has been a UV coated 
Thompson CCD with 1024$\times$1024 square pixels of 19$\mu$m size.
The GAIA spectral range is covered without gaps in a single order by the
Asiago Echelle spectrograph. The actual observations however extended over a
much larger wavelength interval (4550-9600 \AA). Here we will limit the
analysis to the GAIA spectral interval; the remaining, much larger
wavelength domain will be analyzed elsewhere together with devoted
multi-band photometry from ground based observations. The spectra
have been extracted and calibrated in a standard fashion using IRAF
software packages running on a PC under the Linux operating system. The high
stability of the wavelength scale of the Asiago Echelle spectrograph has
been discussed in Paper I. The results of radial velocity measurements are 
given in Table~3.

\section{Modeling}

We use an upgrade of the setup described in Paper I. The binary modeling code 
(Wilson 1998) was combined with Van Hamme's limb darkening coefficients 
(van Hamme 1993), a fitting package, a graphical user interface 
and utilities like reddening corrections to form PHOEBE (Pr\v{s}a 2003). 
The package is able to run on any Unix platform.
It may constitute the first step toward automated solution-finding routines 
that will be needed to interpret the vast number of binary systems to be 
observed by GAIA. All results were independently derived also by the 
WD98K93 code (Milone et al.\ 1992) and WD2002 code (Kallrath et al.\ 1998)
that are briefly described in Paper~I. We found that the results 
are in agreement. 

The usual approach to binary star modeling is to use only relative 
photometry obtained in each filter. Depths of eclipses in different filters 
constrain the ratio of the stellar temperatures, while the absolute 
temperature scale is tuned by judging the primary star 
temperature from the system colour. 

In our case both stars are of similar brightness and the light curves 
are quite noisy. This requires some modifications to the usual approach.
Hipparcos observed in three filters. The observations obtained in the 
broad band $H_P$ filter have an acceptable accuracy, while those in 
the Tycho experiment's $B_T$ and $V_T$ bands are generally very noisy. 
We use the absolute system colours at quarter phase to fix the absolute 
temperature scale. The transformation between the Tycho and 
Johnson systems is the same as in Paper~I:
\begin{eqnarray}
V_J & = & V_T - 0.090 \times (B-V)_T \\
(B-V)_J & = &0.85 \times (B-V)_T
\end{eqnarray}
Temperatures of both stars are similar, so the temperatures 
of the stars, $T_1$, $T_2$ and their radii $R_1$, $R_2$ are connected to the 
surface-weighted effective temperature of the source at quarter phase 
$T_{1+2}$ by the relation:
\begin{equation}
R_1^2 T_1^4 + R_2^2 T_2^4 = (R_1^2 + R_2^2) T_{1+2}^4
\end{equation}
First the Tycho colour index at quarter phase of the model fits to the 
$B_T$ and $V_T$ light curves was transformed to the Johnson 
system (Eq. (2)) and the effective temperature $T_{1+2}$ was determined. 
Modeling of the better quality $H_P$ band observations yielded the 
temperature ratio and, by use of Eq.\ (3), also the absolute temperatures of the 
two stars. The process was reiterated several times to reach a self-consistent 
solution. 
 
Some colour calibrations proposed recently (Bessell 2000) differ from Eq.\ (2) 
and cause effective temperature offsets of $\sim 100$~K. We will comment on 
the changes of the results if these relations were used in the Discussion. 

\section{Results}

Table~4 quotes the derived system parameters together with their formal errors.
Table~5 compares the derived distances to the astrometric results from 
Hipparcos. The data and the curves from the model solutions are plotted 
in Figs.~1--3. 

We note that model fits are generally acceptable. The differences are chiefly 
due to noise in the data and to some degree due to intrinsic variability of 
the stars. A limited number of epochs and their long timespan make 
modeling of transient phenomena such as stellar spots unfeasible. This will 
generally be also the case with data obtained by GAIA. The results were 
obtained assuming the stars are co-rotating. Next we discuss in turn 
the results for each of the objects.

\subsection{V781 Tau}

V781 Tau is an overcontact binary with different primary and secondary temperatures. 
Light curve modeling fixes the quarter phase magnitudes 
($\Phi = 0.75$) to $V_T = 8.56$ and $B_T = 9.16$. This corresponds to the colour 
index $(B-V)_T = 0.60$ or $(B-V)_J = 0.51$ (Eq.\ (2)) which gives $T_{1+2} = 6240$~K.
This result was used to constrain the temperatures of the two stars through Eq.~(3). 
Note that the magnitudes quoted in the Hipparcos catalogue (Table~1) would 
give somewhat different colours. However these magnitudes are just a suitable 
mean of all observations, also the ones close to the photometric eclipses.
Therefore it is correct to use the quarter phase light curve fit and not the 
mean colours. 

Spectroscopic observations determine absolute size of the system 
and individual masses as a function of the system inclination. A detailed 
reflection treatment was used to compute the photometric curves. The $H_P$ light 
curve constrains relative sizes and temperature ratio of both stars. 
We found the system is actually filling 
its Roche lobes up to the $L_1$ point. The stars are of unequal temperature 
($T_1 - T_2 \sim 220$~K). This difference was explained by mass transfer between 
the stars and the corresponding gravitational energy release 
(Liu \&\ Yang 2000). A small period decrease ($dP/P = -5.0 \times 10^{-11}$) 
was also claimed to be an effect of mass transfer. 
We note that any mass lost from the system through the $L_2$
point would carry away roughly twice the mean value of the specific angular momentum. 
Mass loss through the $L_2$ point can therefore decrease the total angular momentum of 
the system, so it may be partially responsible for shortening of the orbital period. 
The value of the time derivative is too small to be detectable from  
data used in this study. 

Lu (1993) published a spectrophotometric study roughly at the same accuracy level 
as reported in Table~4. The values of individual parameters are generally consistent,
with some differences possibly arising from the simplified software he used for 
modeling. In particular he adopted lower effective temperatures ($T_{1,2} = $5950, 
5861~K) but with a large error bar of 200~K. Therefore the system in his analysis
turns out to be fainter and at a smaller distance (72~pc). 

We note that the formal error bars on temperatures as given by the WD98 code
can be increased due to systematic effects.  True uncertainty can reach 100~K, 
increasing the uncertainty on the distance (Table~5) to 4.5\%\ or 3.6~pc. Temperatures 
of both stars may be also influenced by reddening. 
V~781 Tau lies just $0.2^\circ$ from the Galactic plane. One may expect $E(B-V) = 0.09$, 
and $A_V = 0.3$~mag (Perry and Johnston 1982). In our calibration the effective 
temperature $T_{1+2}$ would raise to 6540~K and the bolometric magnitude of 
the system would be brighter by 0.33~mag. Note that this brightening almost 
cancels out with the value of the total extinction. So reddening has little 
influence on the distance of the system reported in Table~5.

\subsection{UV Leo}

UV Leo is a close binary with a pronounced spot activity that is expected 
to be common between G/K type binaries to be observed by GAIA. The spots 
cause vertical offsets in the brightness of the object on a time-scale 
of weeks to months (Miku\v{z} et al.\ 2002). Such intrinsic 
variability may be contributing to the scatter of H$_P$ observations in Fig.\ 2. 
Magnetic activity may be also responsible for part of the scatter of 
the radial velocity curves ($\sigma_{RV1,2} = 17.6 $~km~s$^{-1}$, Table 4). In fact the Ca~II lines from the secondary 
on JD~2451896 show hints of multi-component profiles, typical for spotted  
stars. This structure, though below the level suitable for detailed 
analysis in our (and usually also GAIA's) coverage of the Ca~II lines, 
obviously increases the scatter of derived radial velocities. 

The fits to the V$_T$ and B$_T$ curves
give a quarter phase colour $(B-V)_T = 0.72$, corresponding to $(B-V)_J = 0.61$.
This is consistent with the colours derived by Popper (1997). 
For main sequence stars this colour index translates into $T_{1+2} = 5900$~K. 
This constraint was adopted during our spectrophotometric model fitting.

Popper (1997) published a spectrophotometric solution deriving the average
masses, radii and temperatures of both stars. Here we derive the parameters 
also for individual stars. The results are generally consistent.

\subsection{GK Dra}

Similar to UV~Leo, GK~Dra also features intrinsic variability of its components.
The variability is however not caused by spots but by a likely $\delta$-Sct 
variability on the secondary star (Dallaporta et al.\ 2002). This 
variability has an amplitude of $\sim 0.05$~mag, so it is partially responsible 
for the scatter in the H$_P$ curve in Fig.\ 3. The V$_T$ and B$_T$ curves are 
very noisy. Still they provide an average quarter phase colour $(B-V)_T = 0.39$, 
corresponding to $(B-V)_J = 0.33$ and effective temperature 
$T_{1+2} = 7000$~K. The photometry to be obtained by GAIA will be 
of much higher accuracy ($\sigma \sim 0.001$~mag) than Tycho observations. 
This will provide for accurate colour information also during eclipses and 
therefore constrain the temperature of either star. 

Hipparcos catalogue lists an orbital period of 16.96~days. Dallaporta et al.\ 
(2002) showed by a devoted ground-based observation campaign that the true 
period is 9.97 days. The error in the Hipparcos results can be 
traced to the fact that the orbital period had to be derived from only 124 
points. The system is detached so only 15 point fell into either eclipse. 
Spectroscopic information obtained by GAIA will greatly alleviate such 
problems (see Zwitter 2003 for detailed simulations). This is a consequence 
of the fact that every radial velocity point contributes to period determination 
and not only those falling into eclipses as for photometric observations.

\section{Discussion}

Our analysis used the Tycho to Johnson colour transformation from the 
Hipparcos catalogue as given in Eqs.\ (1) and (2). The magnitude 
measurements themselves were obtained from the Hipparcos and Tycho-1 epoch 
photometry as available through the CDS. Recently Bessell (2000) published 
modified calibrations that would make the $(B-V)_J$ colours redder by  
0.03 to 0.04 mag. The $T_{1+2}$ temperatures for V781 Tau, UV Leo and 
GK Dra would be lower for 120~K, 100~K and 160~K, respectively. A modified 
$V_J-V_T$ vs.\ $(B-V)_T$ relation would also make their apparent $V_J$ 
magnitudes $\sim 0.01$~mag brighter. Such small corrections cannot significantly 
modify the limb-darkening and other coefficients that depend on the absolute 
value of the temperature. But they do change the bolometric magnitudes and so 
distances. In our case the absolute bolometric magnitudes for V781 Tau, 
UV Leo and GK Dra would be 0.28, 0.19 and 0.46~mag fainter and the derived 
distances 14, 9 and 23 \%\ larger. The issue of absolute colour calibrations 
of the Tycho passbands does not seem to be a closed one. The new version of 
the Tycho catalogue (Tycho-2) quotes the old calibration (Eq.\ (2)) again. 
We therefore prefer to remain with the same calibration as used in Paper~I 
with the possible modifications clearly spelled out. 

\section{Conclusions}

The paper clearly demonstrates the potential of GAIA to derive accurate 
orbital solutions even for stars with intrinsic variability or for contact 
cases. GAIA will observe any object only around a hundred times. 
This will complicate the determination of orbital period of wide detached 
systems. Spectroscopic information will be particularly useful to determine 
the orbital period in such cases and also for a vast majority of binaries 
which are non-eclipsing. Spectroscopic information can be used also to 
derive orbital eccentricity as demonstrated by GK~Dra.   

Absolute scale of the system provided by spectroscopic orbit can be used 
to derive masses and sizes of the system components at a 1-2\%\ level  
(Table~4). So these stars can be absolutely placed 
on an H-R diagram. Exact coevality of both stars in a binary make  
for a useful study of stellar isochrones. Munari (2003) 
discusses how additional information, like metallicity, will be obtained 
from the GAIA data. 

The distances derived from orbital solutions compete or are superior 
to the Hipparcos astrometric measurements. We note that the present analysis 
may be influenced by uncertain calibrations in the noisy photometry obtained 
from the Hipparcos Tycho experiment. But for the case of GAIA the errors 
quoted in Table~5 are realistic, as the stellar temperatures and 
reddening will be known with high precision from a multi-band photometry. Note 
also that measurement of distances from orbital solutions, especially for 
overcontact binaries, is limited only by relative faintness of the objects
at large distances. So hot contact binaries will be a useful tool to 
gauge distance throughout the Galaxy and beyond.

GAIA will be able to detect also intrinsic variability of binary components. 
Degree of derivable physical information depends on the nature of the 
variability. Stellar spots will be very common but difficult to describe. 
These are transient phenomena, so the star will look different on each 
of the 100 transits during the 5-yr mission lifetime. This can be seen also in 
our data. Different levels of quarter phase maxima in the V781~Tau light 
curve (Cereda et al.\ 1988) were used to claim the presence of polar spots
(Lu 1993). But Hipparcos light curves do not reveal such details. Also 
UV Leo is an object with occasional spots that change the overall 
system brightness. The fact that we ignored such phenomena but still derived 
quite accurate orbital solutions in two systems suggests that magnetic phenomena cannot 
jeopardize the derivation of binary star parameters to some limit of accuracy. 
Other types of variability, like $\delta$-Sct variability in GK~Dra (Dallaporta 
et al.\ 2002, Zwitter 2003) maintain its phase, so they will be easily detectable
from GAIA data. Orbital period changes, e.g.\ due to passages of the third 
body will be quite uncommon and difficult to detect due to a limited mission 
lifetime. 

This work reassures us of the high quality of physical information 
recoverable from GAIA's observations of eclipsing binaries. In future papers 
of this series we plan to explore more objects with intrinsic variability 
as well as some 
double lined systems with triple components.

\begin{table}[!t]
\tabcolsep 0.08truecm
\caption{Comparison between the Hipparcos distances and those derived from the 
parameters of the modeling solution in Table~4. Only formal errors quoted 
in Table~4 were taken into account. As explained in the text the actual 
distances may have a bit larger uncertainties.}
\begin{tabular}{lcc}
\hline
          &Hipparcos             & this paper      \\
          & (pc)                 &  (pc)           \\
          &                      &                 \\
V781~Tau  & 81$^{91}_{73}$       & 81$\pm$1.0    \\ 
          &                      &                 \\
UV~Leo    & 91$^{103}_{83}$      &  92$\pm$6      \\
          &                      &                 \\
GK~Dra    &297$^{373}_{246}$     &  313$\pm$14     \\
          &                      &                 \\
\hline
\end{tabular}
\end{table}

\acknowledgements{Generous allocation of observing time with the 
Asiago telescopes has been vital to the present project. We would like 
to thank Bob Wilson who continues to provide us with the last versions 
of the Wilson-Devinney program and for advice on its usage. EFM acknowledges 
Josef Kallrath's help in further developments of spot analysis with 
the WD2002 code. The financial 
support from the Slovenian Ministry for Education, Science and Sports (to TZ and
AP), of the NSERC and University of Calgary Research Grants Committee (EFM),
and through the Polish KBN Grant No.5 P03D 003 20 (TT) is kindly acknowledged.}
\\

\end{document}